\documentclass[reprint,prl,twocolumn,showpacs,superscriptaddress,aps,longbibliography]{revtex4-1}

\usepackage[utf8]{inputenc}
\usepackage[T1]{fontenc}

\usepackage{graphicx}
\DeclareGraphicsExtensions{.png,.jpg,.eps,.pdf}

\usepackage{float}
\usepackage{xcolor}
\usepackage{amsmath}
\usepackage{float}
\usepackage[colorlinks=true,citecolor=blue]{hyperref}

\begin{document}

\title{Nature of the unconventional heavy fermion Kondo state in monolayer CeSiI}

\author{Adolfo O. Fumega}
\affiliation{Department of Applied Physics, Aalto University, 02150 Espoo, Finland}

\author{Jose L. Lado}
\affiliation{Department of Applied Physics, Aalto University, 02150 Espoo, Finland}

\begin{abstract}
CeSiI has been recently isolated in the ultrathin limit, establishing CeSiI as
the first intrinsic two-dimensional van der Waals
heavy-fermion material up to 85 K. 
We show that, due to the strong Ce spin-orbit coupling, the local moments develop a multipolar real-space magnetic texture, leading to local pseudospins with a nearly vanishing net moment. To elucidate its Kondo-screened regime, we extract from first principles the parameters of the Kondo lattice model describing this material. We develop a fractional pseudofermion methodology in combination with ab initio calculations to reveal the nature of the Kondo-screened heavy fermion state in CeSiI. Using this formalism, we analyze the competing magnetic interactions leading to a heavy-fermion order
as a function of the magnetic exchange between the localized f-electrons and the strength of the Kondo coupling. Our results show that the magnetic exchange interactions promote an unconventional momentum-dependent
heavy-fermion Kondo screened phase, establishing the nature of the heavy-fermion behavior of ultrathin
CeSiI observed experimentally.

\end{abstract}

\maketitle

The coexistence of electronic orders in two-dimensional (2D) materials establishes a rich platform for the emergence of new physics. Since the isolation of van der Waals monolayers, a variety of magnetic orders have been observed in the 2D limit, including ferromagnetic\cite{Huang2017,Fei2018}, quantum spin-liquid candidates\cite{Ruan2021,Chen2022}, and multiferroic order\cite{Song2022,2023arXiv230911217A}. 
Furthermore, exploring the unique degrees of freedom of van der Waals materials, namely the easy and clean stacking of layers forming heterostructures and introducing twist angles between layers, has allowed the emergence of new magnetic orders, including orbital ferromagnets\cite{Serlin2020,Park2023}, and heavy fermion Kondo lattice materials\cite{Vao2021,PhysRevLett.127.026401,PhysRevLett.129.047601,Zhao2023,PhysRevB.106.L041116,Jang2022,Guerci2023,zhao2023emergence}.
The recent isolation of  CeSiI\cite{Posey2024} in the ultra-thin
limit establishes heavy-fermion Kondo insulators as a new member in the family of van der Waals building blocks.

\begin{figure}[t!]
\includegraphics[width=\columnwidth]{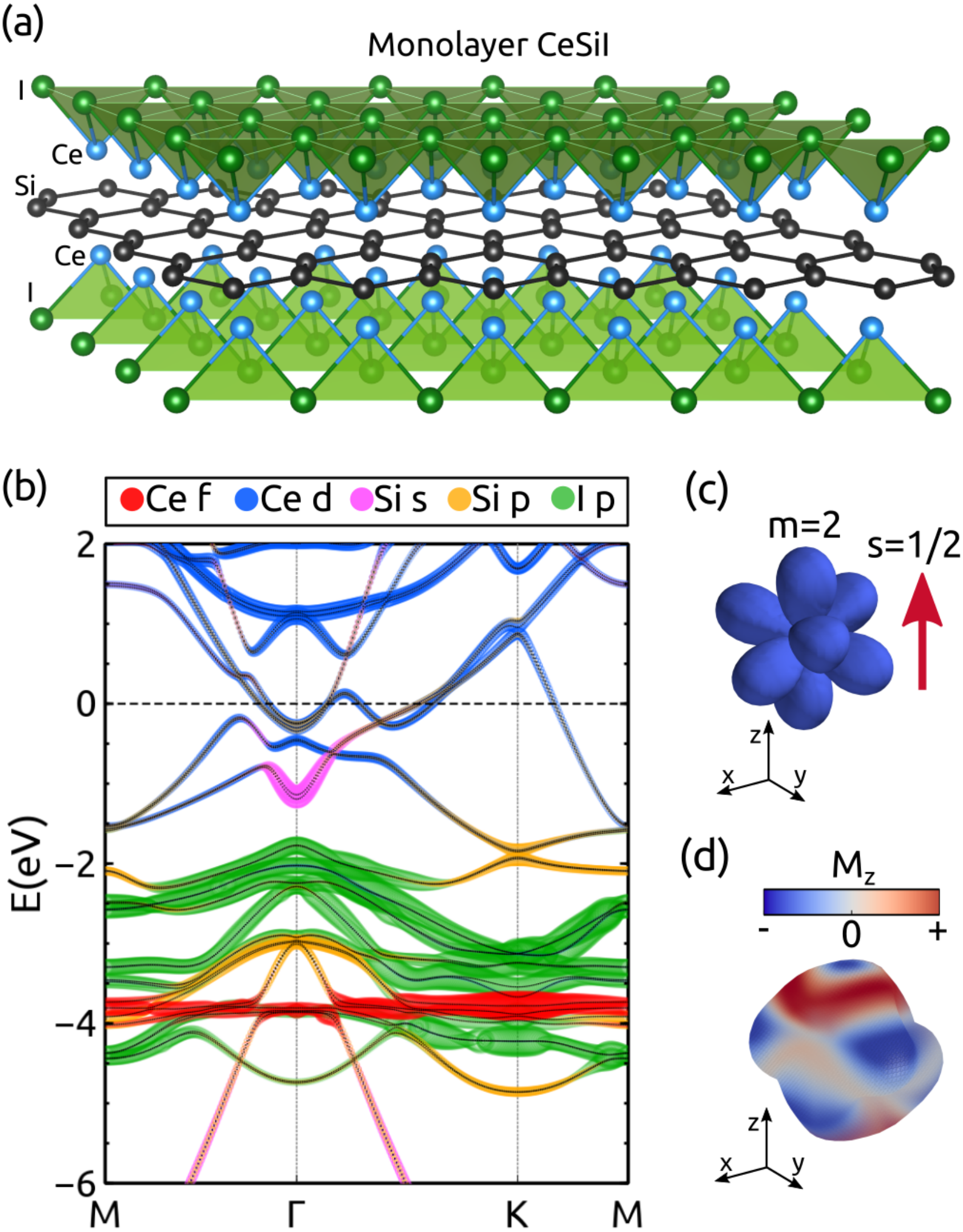}
\caption{(a) Structure of monolayer CeSiI.
(b) DFT orbital-resolved band structure of
CeSiI, enforcing a symmetry-broken ferromagnetic arrangement of Ce magnetic moments. 
Panels (c) and (d) show the magnetization of the occupied f-state in the absence (c) and presence (d) of spin-orbit coupling.
In the latter case, the net magnetization of the occupied orbital is quenched, leading to a microscopic spin texture in the occupied state.
}
\label{CeSiI_struct}
\end{figure}

Heavy-fermion systems emerge due to the coexistence of a magnetic lattice that is coupled to a nearly free 2D electron gas forming what is known as a Kondo lattice.\cite{2015arXiv150905769C} 
Traditionally, these two ingredients have been found in bulk rare-earth compounds, which bring together magnetic moments from the localized f orbitals and an electron gas from the delocalized orbitals.\cite{RevModPhys.56.755}
These Kondo systems display intriguing phase diagrams in which unconventional superconductivity, quantum critical phases, or the already-mentioned heavy-fermion order have been reported.\cite{doi:10.1126/science.1191195,2015arXiv150905769C} Therefore, identifying new heavy-fermion systems is proven to be a powerful strategy for studying novel exotic phenomena.

In the realm of van der Waals materials, heavy-fermion systems have been artificially engineered in heterostructures with different 2D materials including 1T-1H TaS$_2$ bilayers, MoTe$_2$/WSe$_2$ bilayers and MoS$_2$ bilayers.\cite{Vao2021,Zhao2023,Guerci2023}
Now, monolayer CeSiI brings Kondo physics to a single van der Waals block, allowing us to study these systems from a novel perspective combining typical surface-science experimental techniques such as scanning tunneling microscopy\cite{RevModPhys.92.011002} and exploiting the degrees of freedom characteristic of van der Waals materials.\cite{vdwHT2013}
However, theoretical studies on monolayer CeSiI are still scarce\cite{Jang2022}
due to the difficult treatment of Kondo systems from an \emph{ab initio} perspective. This obstacle hinders the study of the complex phase diagram that could arise in CeSiI van der Waals heterostructures.

In this work, we analyze the emergent heavy-fermion phase in monolayer CeSiI. We introduce a formalism based on Density Functional Theory (DFT) calculations combined with an auxiliary pseudofermion variational method. This approach that we termed DFT+pseudofermion allows us to study the heavy-fermion order emerging in this compound. We show that the strong spin-orbit coupling effects lead to an exotic magnetic texture in the local Ce moments, leading to an emergent pseudo-spin Kondo lattice. Using the DFT+pseudofermion formalism, we capture the first principles multiorbital electronic structure together with the many-body Kondo screening that occurs in CeSiI. In particular, we establish that the competition between the Kondo coupling and the magnetic exchange interactions between the localized Ce f-electrons gives rise to the unconventional nodal heavy-fermion order that can be proved experimentally.

It is instructive to start analyzing the electronic structure of the monolayer CeSiI arising directly from DFT. The structure of monolayer CeSiI is shown in Fig. \ref{CeSiI_struct}a, where
two triangular lattices of Ce and I atoms encapsulate a staggered honeycomb lattice of Si atoms forming the single van der Waals block.
Fig. \ref{CeSiI_struct}b shows the orbital-resolved band structure of monolayer CeSiI obtained from DFT calculations using the LDA+U formalism enforcing a symmetry-broken ferromagnetic arrangement and including spin-orbit coupling (SOC), with U corresponding to the on-site Coulomb interaction of the localized Ce f-electrons. In the plot, a U=8 eV was considered for the f electrons. However, the results and discussion presented here are not qualitatively affected by the election of a different U in the strongly localized limit. 
The strong electron-electron interactions in the f-orbitals of Ce give rise to the formation of a local magnetic moment, leading to a set of deep occupied nearly-dispersionless flat bands (shown in red in Fig. \ref{CeSiI_struct}b).
The electrons belonging to Si sp orbitals and  Ce d orbitals are in comparison strongly dispersive, leading to a metallic behavior (shown in pink, yellow, and blue respectively in Fig. \ref{CeSiI_struct}b). 
These conduction states show strong hybridization between the different elements. Enforcing a ferromagnetic arrangement between the f-electrons is a technical consideration that allows a simpler treatment in the single unit cell and has no major effect on the analysis of the orbital character of the bands.

\begin{figure}[t!]
\begin{center}
\includegraphics[width=\columnwidth,draft=false]{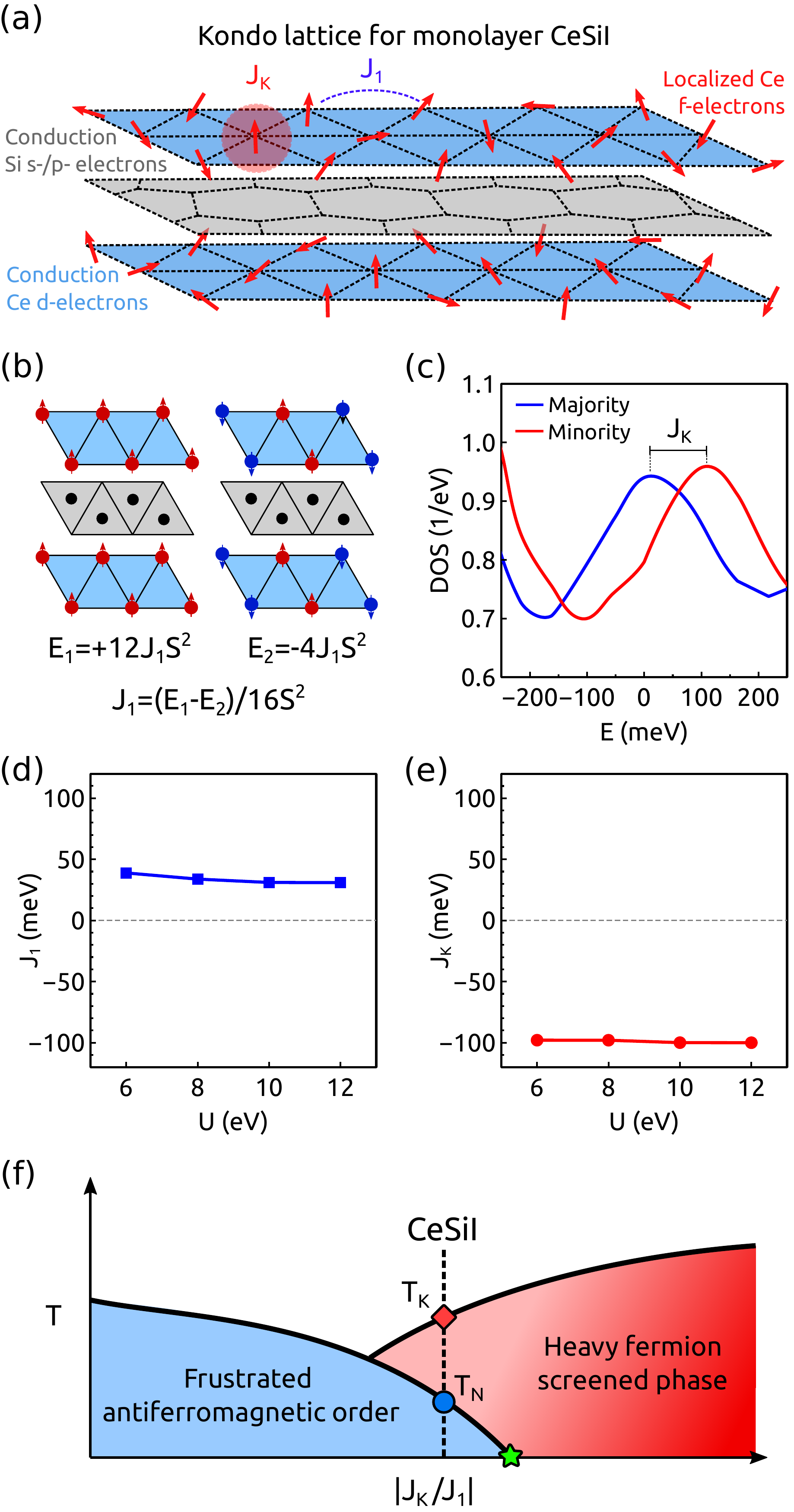}
\caption{(a) Schematic low-energy model for CeSiI. The spin lattice stems from the localized f-electrons of Ce, while the conduction electrons correspond mainly to the d-electrons of Ce. 
First neighbors spins are coupled via a magnetic exchange interaction $J_1$, and Kondo coupled to the conduction electrons with $J_K$.
(b) Magnetic configurations used to estimate  $J_1$. (c) Spin-polarized density of states plot used to estimate $J_K$.
Evolution of $J_1$ (d) and $J_K$ (e) from DFT+U as a function of the Coulomb interaction U of the f-orbitals.
(f) Schematic heavy fermion phase diagram of the Kondo lattice model, where
CeSiI is found to enter into a heavy-fermion phase below T$_K$ and below T$_N$ into an antiferromagnetic one. The green star denotes the quantum critical region. 
}
\label{Fig_DFT_parameters}
\end{center}
\end{figure}

A key feature of this material is the existence of a very strong spin-orbit coupling. This strong spin-orbit coupling has a major effect on the local moment of the f-orbitals. 
In the absence of spin-orbit coupling, the f-electrons display an orbital momentum m=2 and spin 1/2 (Fig. \ref{CeSiI_struct}c). \footnote{The band structure in the absence of spin-orbit coupling is shown in the Supplemental Information.}
However, when spin-orbit interactions are included, the original net $S=1/2$ moment transforms into a complex spin structure in the f-manifold, with a quenched net magnetic moment  (Fig. \ref{CeSiI_struct}d). 
This orbital degeneracy together with strong spin-orbit coupling effects can give rise to a hidden magnetic order\cite{PhysRevLett.103.107202,PhysRevB.62.4880,Okazaki2011,PhysRevLett.81.3723,RevModPhys.83.1301,Mydosh2020}, characterized by a multipolar localized spin texture with a quenched magnetic moment that is difficult to identify in experiments. 
From an effective model point of view,
the unpaired electron behaves as a pseudo-spin $1/2$, which is the required feature for the emergence of heavy-fermion phenomena, and monolayer CeSiI stems from the
Kondo lattice Hamiltonian\cite{Kasuya1956,RevModPhys.69.809}

\begin{equation}
\label{eq_kondo_hamiltonian}
\begin{aligned}
    H = \sum_{ij}
    t_{ij} c^\dagger_i c_j
    + \sum_{\langle i,\alpha\rangle} J_K \vec S_\alpha \cdot \vec \sigma_{s,s'} c^\dagger_{i,s} c_{i,s'}
 \\   + J_1 \sum_{\langle \alpha,\beta \rangle} \vec S_\alpha \cdot \vec S_\beta
    \end{aligned}
\end{equation}

where $c^\dagger_i$ is the creation operator for Wannier conduction band electrons, $t_{ij}$ corresponds to their hopping energy, $\langle \rangle$ denotes first neighbors and $J_1$ is the exchange coupling between first-neighbor pseudo-spin sites of the f electrons, and $J_K$ is the Kondo coupling between the localized pseudo spins and the conduction electrons. 
It is important to note that in the Kondo problem, these pseudo spins act as a real spin 1/2, since the Hilbert subspace representation of these pseudo spins corresponds to one of spins 1/2 despite the local magnetic texture that they might display. Therefore, the J$_1$ interaction captures the magnetic exchange interaction between these pseudo spins.
The schematic of this model is shown in  Fig. \ref{Fig_DFT_parameters}a. 
The previous Hamiltonian realizes a Kondo lattice model, whose physics is controlled by two main parameters.
The exchange coupling $J_1$ promotes magnetic ordering in the magnetic lattice formed by the f-electrons. In contrast, the Kondo coupling $J_K$ promotes the screening of each spin site, by forming a singlet with a conduction electron of the conduction gas\cite{RevModPhys.69.809}. This screening of spin sites in the lattice creates a coherent set of scattering centers, leading to the appearance of a heavy fermion gap\cite{Riseborough2000}. The competition between those two energy scales dominates the physics of a heavy-fermion monolayer.

Traditional DFT calculations cannot directly capture the many-body heavy-fermion behavior in the electronic structure. However, they can be used to provide an estimate of the competing parameters entering the Kondo lattice Hamiltonian. 
Specifically, the exchange coupling can be extracted by comparing the energies between different magnetic arrangements (Fig. \ref{Fig_DFT_parameters}b).
It is important to point out that the exchange coupling $J_1$ can have both direct and, indirect exchange through the I atoms, and an indirect interaction mediated by the electron gas. This last interaction, known as RKKY interaction, is the dominant one due to the strongly localized f-electrons residing within the electron gas. First principle methods incorporate all the contributions, and therefore the exchange extracted represents the net one.
On the other hand, the Kondo coupling stems from the induced spin splitting in the conduction bath by the local magnetic moment. Therefore, $J_K$ can be estimated as the energy shift in the Density of States (DOS) of the conduction bands at the Fermi level (Fig. \ref{Fig_DFT_parameters}c).
The DFT+U estimation of $J_1$ and $J_K$ is shown in Figs. \ref{Fig_DFT_parameters}d and  \ref{Fig_DFT_parameters}e respectively. We can observe a robust trend in the estimation of these quantities as a function of the onsite Coulomb interaction of the f-electrons U. An antiferromagnetic $J_1$ is obtained in good agreement with magnetometry measurements\cite{PhysRevMaterials.5.L121401} which suggests that the phase diagram of the competing $J_K$-$J_1$ triangular Kondo lattice (eq. (\ref{eq_kondo_hamiltonian})) would display a quantum critical phase between a frustrated magnetic order and the heavy fermion order\cite{Coleman2010,Tokiwa2015,Zhao2019,Ramires2019,Kavai2021}, instead of the usual quantum critical point between a ferromagnetic and the heavy-fermion order (Fig. \ref{Fig_DFT_parameters}d). DFT+U provides an estimation of $J_1\simeq 30$ meV and $J_K\simeq -100$ meV, which suggests that monolayer CeSiI can enter the heavy-fermion phase dominated by $J_K$, but with a non-negligible magnetic exchange $J_1$ between the f electrons. This finite $J_1$ exchange promotes the formation of a frustrated
magnetic order at the lowest temperature below 7.5 K,
with screening from Kondo physics dominating between 7.5 K and
85 K\cite{Posey2024}.

The screened coherent heavy fermion
Kondo regime\cite{Posey2024} can be
directly accounted
by the Kondo lattice model.
The previous Kondo lattice model can be solved using an auxiliary fermion  (pseudofermion or parton) formalism for Kondo sites\cite{PhysRevB.29.3035,PhysRevLett.57.877,PhysRevB.35.3394,PhysRevB.93.035120,PhysRevLett.112.116405,PhysRevB.35.5072}
$\vec S_\alpha = \frac{1}{2} \sum_{s,s'} \sigma_{s,s'} f^\dagger_{s,\alpha} f_{s',\alpha}$ with the Fock contraint $\sum _s f^\dagger_{s,\alpha} f_{s',\alpha} = 1$.
By making the replacement of spin operators by auxiliary fermions, the Hamiltonian becomes biquadratic in field operators.
We can perform a decoupling of the biquadratic term
by introducing the Kondo hybridization functions,
leading to the following effective Hamiltonian

\begin{equation}
\label{eq_Ham_pseudo}
\begin{aligned}
    H & = \sum_{\nu,\mathbf k}
    \epsilon_\nu (\mathbf k) c^\dagger_{\mathbf k,\nu} c_{\mathbf k,\nu}
    + \sum_{\alpha,\mathbf k} \gamma_1 (\mathbf k)  f^\dagger_{\mathbf k,\alpha} f_{\mathbf k,\alpha}   
     \\
    & + \sum_{\nu,\alpha,\mathbf k} \gamma^{\nu,\alpha}_K (\mathbf k) 
    f^\dagger_{\mathbf k,\alpha} c_{\mathbf k,\nu}
\end{aligned}
\end{equation}

where $f^\dagger_{\alpha,s}$ are the pseudofermions creation operators in reciprocal space. The Kondo hybridization function depends on the Kondo coupling as
$\gamma_K (\mathbf k) \sim J_K \langle c^\dagger_{\mathbf k} f_{\mathbf k} \rangle$ and the dispersion of the pseudo fermions is given by the Fourier transform
of the pseudo-fermion mean-field 
$|\gamma_1 (\mathbf k) | \sim J_1$.
This Hamiltonian can be solved in combination with the first-principles electronic calculations for monolayer CeSiI. A DFT+pseudofermion formalism can be established by expanding the Hilbert space of the DFT states (first term in eq. (\ref{eq_Ham_pseudo})) to include the pseudofermions (second term in eq. (\ref{eq_Ham_pseudo})) and their hybridization with the DFT states (third term in eq. (\ref{eq_Ham_pseudo})). This allows us to compute the electronic structure of monolayer CeSiI in the presence of Kondo screening. Since the unit cell of monolayer CeSiI has two Ce atoms that give rise to two localized pseudo-spins 1/2, four pseudofermions are considered in this DFT+pseudofermion formalism.
In the case of two 1/2-real spins, one would also consider four pseudofermions in the DFT + pseudofermion formalism. The pseudo-spins are a result of the spin-orbit coupling leading to a complex spin texture.
Two channels in each pseudo-fermion do not correspond to up and down, but the two channels of a Kramers pair. Specifically, the two channels of a Kramers pair have the property that one channel is the time-reversal of the other. These two channels are analogous to the two degrees of freedom of an electronic structure featuring strong spin-orbit coupling effects, where the spin degeneracy is associated with two Kramers pairs instead of two pure spin channels. Therefore, both have the same technical implementation in our formalism.
The hybridization function $\gamma_K$ is in general momentum, frequency, and
band-dependent.
The frequency and band dependence 
is incorporated with an ansatz promoting
hybridization with the closest four bands to the Fermi energy
including a frequency-dependent envelop
$e^{-(\epsilon_\nu (\mathbf k)-E_F)^2/J_K^2}$,
that accounts for the decreased hybridization
away from the chemical potential $E_F$.

\begin{figure}[t!]
\begin{center}
\includegraphics[width=\columnwidth,draft=false]{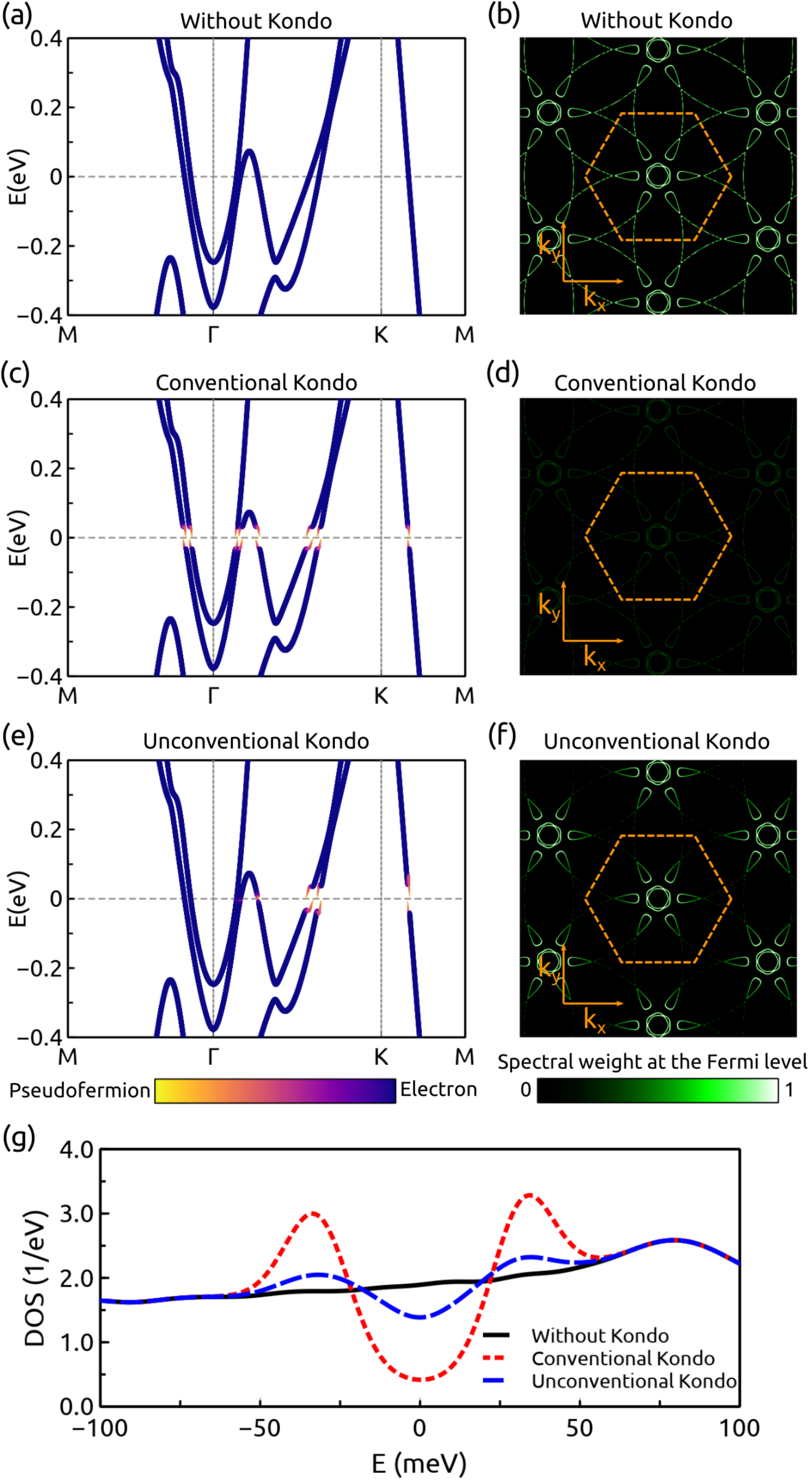}
\caption{
Electronic structure without Kondo hybridization (a,b), with conventional Kondo hybridization (b,c)
and with unconventional nodal heavy-fermion hybridization (d,e).
Panels (a,c,e) show the momentum-resolved spectral function, and panels (b,d,f) show the Fermi surface reconstruction. (g)
Spectral function observed showing the differences between conventional and unconventional nodal hybridization.
}
\label{Fig_kondosolutions}
\end{center}
\end{figure}

\begin{figure*}[t!]
\begin{center}
\includegraphics[width=\textwidth,draft=false]{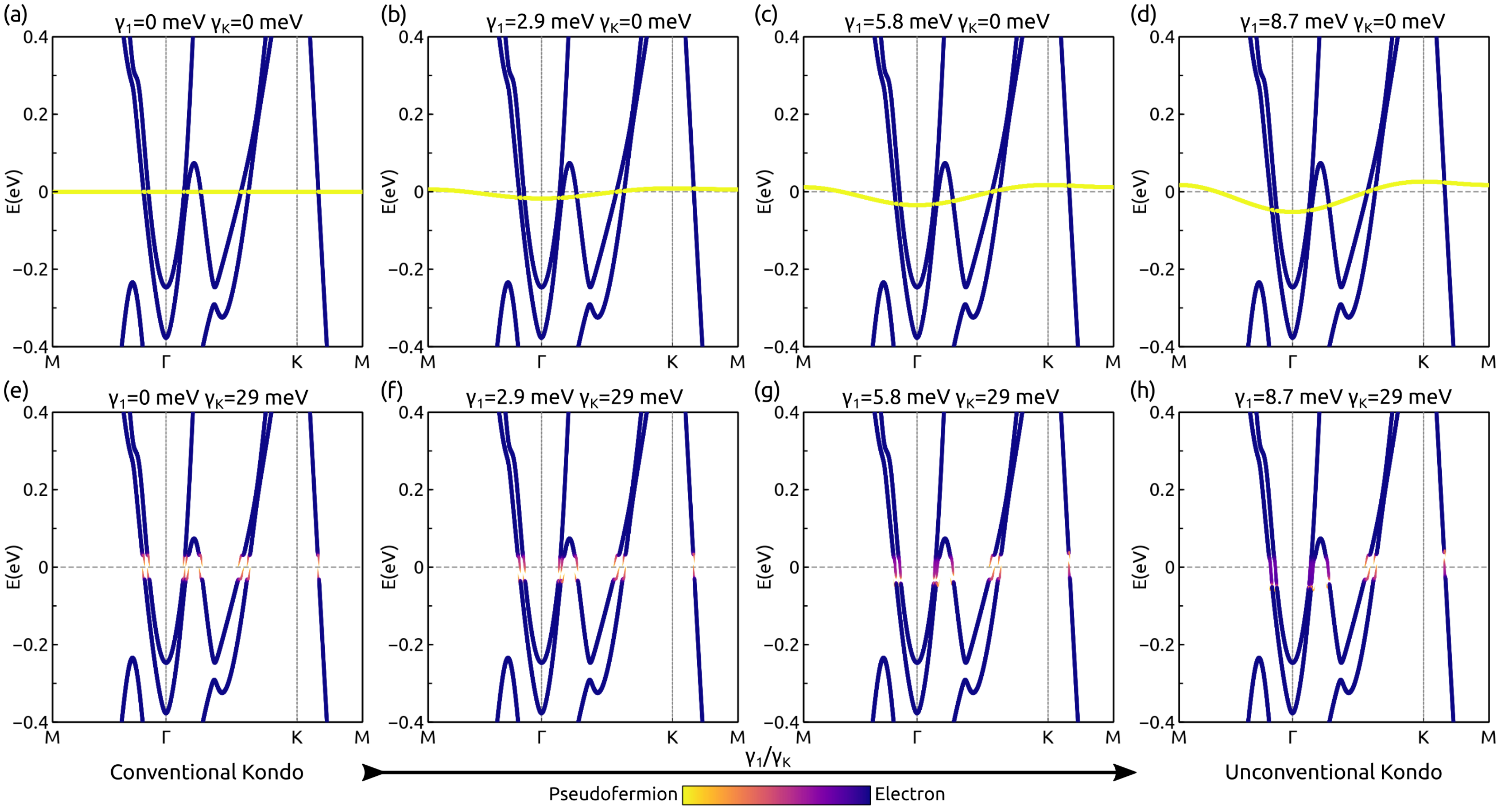}
\caption{Electron-pseudofermion dispersion in the presence of
a sizable exchange coupling $J_1$ between Kondo sites.
Panels (a-d) show the decoupled pseudofermion and electronic structures as reference.
Panels (e-h) show the composite electronic structure in the presence
of Kondo hybridization, showing that the increase of an exchange coupling and
associated pseudofermion dispersion leads to a soft heavy-fermion gap.
}
\label{Fig_J1JK_unconventional}
\end{center}
\end{figure*}

The results for the calculations of monolayer CeSiI in the DFT+pseudofermion formalism
in the non-magnetic Kondo screened regime are summarized in Fig. \ref{Fig_kondosolutions}.
In the absence of Kondo hybridization $\gamma_K (\mathbf k)=0$, the DFT band structure (Fig. \ref{Fig_kondosolutions}a) shows a complex Fermi surface with different pockets around the $\Gamma$ and K points (Fig. \ref{Fig_kondosolutions}b). 
When the Kondo hybridization is finite $\gamma_K (\mathbf k)\neq 0$, gaps appear in the electronic structure at the Fermi energy. From a symmetry point of view, two solutions can emerge when solving the variational Hamiltonian of eq. (\ref{eq_Ham_pseudo}):
i) a conventional s-wave Kondo hybridization opening a gap in the whole Fermi surface (Figs. \ref{Fig_kondosolutions}cd) and with a U-shaped gapped spectral function (Fig. \ref{Fig_kondosolutions}g) or ii) an unconventional nodal f-wave Kondo hybridization opening gaps only around the K points in the Fermi surface (Figs. \ref{Fig_kondosolutions}ef) and displaying a V-shaped spectral function (Fig. \ref{Fig_kondosolutions}g). 
This gapless heavy fermion behavior produces a decreased slope of the band structure (effective electron mass enhancement) mostly around the K points.
This kind of momentum dependence of the Kondo hybridization has already been experimentally observed and analyzed in other heavy-fermion compounds\cite{PhysRevB.75.045109} and in particular Ce-based systems.\cite{PhysRevB.72.033104,PhysRevB.77.155128,PhysRevB.73.224517,PhysRevB.75.054523,PhysRevB.89.115122,PhysRevB.104.125104}  As in the case of CeSiI, they show a k-dependent Kondo hybridization. However, these compounds have a non-van-der-Waals tetragonal structure. Therefore, their nodal behavior displays a different symmetry than the one observed in the triangular lattice of CeSiI. This would have implications in the symmetry of the superconducting order parameters that might occur in the phase diagram of these heavy-fermion systems.\cite{PhysRevB.77.245108}
The nodal behavior could be detected via ARPES experiments as previously done for the other Ce-based compounds or using a Scanning Tunneling Microscope as it has already been used to test unconventional nodal superconductors \cite{Kim2022,Vano_advancedmat_2023}.

The previous classification focused on the case where the Kondo pseudofermions are dispersionless, which is equivalent to considering
vanishingly small exchange coupling between magnetic sites, i.e. $\gamma_1 (\mathbf k)=0$ in eq. (\ref{eq_Ham_pseudo}). Due to the finite exchange $J_1$, the Kondo pseudofermions will develop a finite dispersion ($\gamma_1 (\mathbf k)\neq0$). In the limit of $J_1\gg J_K$ the system will develop magnetic order. Here we will focus on the case $J_1\neq 0$, but below the quantum phase transition to the magnetically ordered regime. The calculations of the electronic structure with DFT+pseudofermion formalism are shown in Fig. \ref{Fig_J1JK_unconventional}, now considering the dispersion in the Kondo spinons. As a reference, we show first how the composite spectra look in the presence of spinon dispersion, but vanishing Kondo hybridization in Figs. \ref{Fig_J1JK_unconventional}abcd for different values of $\gamma_1 (\mathbf k)$. While this situation is not physically observable due to the requirement of Kondo screening to reach the spinon representation, it provides a useful starting point to rationalize the effect of $\gamma_1 (\mathbf k)$ on the conduction gas. The pseudofermion dispersion makes the spinon modes off-resonant with the Fermi surface in major parts of reciprocal space. This effect implies that once the Kondo hybridization is included as shown in Figs. \ref{Fig_J1JK_unconventional}efgh, the gap opening at the Fermi surface becomes much less pronounced, keeping certain parts of the electronic structure gapless.
The dispersive pseudofermion band is closer to the Fermi level and more flat around the K points than around the $\Gamma$ point. Considering that the Kondo hybridization is energy-dependent, i.e. the closer to the Fermi level the pseudofermion energy the stronger the hybridization, this leads to a stronger gap around the K points with respect to the electronic bands around $\Gamma$.
This gapless electronic structure emerges in the presence of a conventional
s-wave Kondo hybridization, emulating an unconventional nodal spectrum. The previous phenomenology shows that the exchange between magnetic sites leads to a weakening of the heavy-fermion gap. In particular, it can be observed that increasing the $\gamma_1/\gamma_K$ ratio promotes the unconventional nodal Kondo hybridization. 
The multiorbital character, meaning having several bands crossing the Fermi level, provides a complex Fermi surface with pockets around the $\Gamma$ and K points (Fig. \ref{Fig_kondosolutions}b). In the presence of Kondo hybridization, the magnetic exchange interactions promote the formation of the Kondo gap around the K points.
DMFT calculations have been performed on bulk CeSiI\cite{Jang2022}. In this analysis, a conventional Kondo peak is captured for this system. In order to study an unconventional nodal behavior within a DMFT methodology a local dependency in the self energy could be included. This would introduce in the DMFT calculations the nodal effect that we are capturing by the pseudofermion dispersion within our auxiliary fermion mean-field theory combined with DFT.
Finally, it is worth noting that below the magnetic transition temperature 7.5K,
the previous non-magnetic fully screen regime gives rise to a spin polarized state
featuring Kondo correlations, and its treatment requires including finite magnetic ordering
in the pseudofermion formalism.

The 2D nature of this material provides promising strategies to tune the ground state. Strain in the monolayer would allow controlling the ratio between $J_K/J_1$,
allowing us to tune the unconventional nature of the heavy fermion order or to push the system towards a quantum phase transition to a magnetically ordered state (Fig. \ref{Fig_DFT_parameters}f). 
In particular, strain can allow to shift the ratio $J_K/J_1$, drifting the system
to heavy fermion regime at $T=0$ and completely suppressing magnetic ordering
even below 7.5K. 
Interestingly, strain will also impact the local crystal field in the Ce atoms, potentially leading to orbital transition in the local magnetic sites. The frustrated nature of the underlying magnetic lattice makes the system ideal to explore the interplay between quantum magnetism and heavy fermion physics. Furthermore, its monolayer limit will potentially allow tuning this heavy-fermion material directly with a gate, allowing to dope the Kondo lattice and drifting the system towards a hidden order or unconventional superconducting state\cite{PhysRevLett.52.679} as observed in other heavy-fermion compounds.  

To summarize, we have presented the microscopic analysis of the heavy fermion state in the van der Waals monolayer CeSiI. Using first-principles methods, we established the existence of local moments in Ce realizing a multipolar internal magnetic texture with vanishingly small local moment. These results show that strong spin-orbit coupling effects render this system into a Kondo lattice system with pseudo-spin $1/2$. Our first-principles methods allow us to extract the relevant energy scales of the Kondo lattice model, including the exchange coupling between the localized moments and the Kondo coupling. We introduce a DFT+pseudofermion formalism which combines the first principles calculations with a parton pseudofermion methodology, allowing us to directly compute the multiorbital heavy-fermion electronic structure. We showed that the sizable exchange coupling between moments together with the multi-band nature of this system leads to a momentum-dependent heavy-fermion hybridization in the Kondo screened regime, in comparison with the stronger gap opening in single band
heavy-fermion systems. Our results establish the first principles electronic structure of CeSiI, exemplifying how a combination of density functional theory and pseudofermion formalism allows the modeling of complex heavy-fermion van der Waals materials.



\textbf{Acknowledgements:}
We acknowledge the computational resources provided by the Aalto Science-IT project, and the financial support from the Academy of Finland Projects No. 331342, No. 336243, and No. 349696 and the Jane and Aatos Erkko Foundation. We thank A. Pasupathy for useful discussions.

\bibliography{CeSiI_monolayer}

\end{document}